%% file: Template.tex
\documentclass{article}
\usepackage{spconf,amsmath,graphicx,hyperref}
\usepackage{makecell}
\usepackage{booktabs}
\usepackage{multirow}
\usepackage{threeparttable}
\usepackage{tabularx}
\usepackage{xcolor}
\usepackage[table]{xcolor}
\usepackage{stfloats}
\usepackage{amsfonts}

\title{Auden-Voice: General-Purpose Voice Encoder for\\ Speech and Language Understanding}

\name{Mingyue Huo$^{1\ast}$\thanks{ $\ast$ This work was done during an internship at Tencent AI Lab, USA.}, Wei-Cheng Tseng$^{2\ast}$, Yiwen Shao$^{3}$, Hao Zhang$^{3}$, Dong Yu$^{3}$\vspace{-4mm}}

\address{
    $^{1}$University of Illinois Urbana-Champaign \quad
    $^{2}$University of Texas at Austin \quad
    $^{3}$Tencent AI Lab, USA \\
    \footnotesize\texttt{mhuo5@illinois.edu, yiwenyshao@global.tencent.com\vspace{-2mm}}
}

\begin{document}
%
\maketitle
\begin{abstract}
Human voice encodes both identity and paralinguistic cues, yet encoders in large audio–language models (LALMs) rarely balance both aspects. In this work, we present a study toward building a general-purpose voice encoder that captures nuanced voice cues. Through a comprehensive evaluation, we find that multi-task training yields the most balanced representations, whereas Contrastive Language–Audio Pretraining (CLAP) primarily improves retrieval without enhancing paralinguistic understanding as expected. Our final encoder, \texttt{Auden-voice}, also demonstrates strong performance when integrated with LLMs. The code and training recipes will be released with the audio understanding toolkit \texttt{Auden}.

\end{abstract}


%
\begin{keywords}
\small
voice encoder, representation learning, multi-task learning, contrastive learning, speech–language models
\vspace{-1.5mm}
\end{keywords}

\input{sections/1_introduction}

\input{sections/2_method}

\input{tables/result_1}

\input{sections/3_evaluation}

\input{sections/4_results}

\input{sections/5_discussion}


\clearpage
{\fontsize{8pt}{7.9pt}\selectfont
\bibliographystyle{IEEEbib}
\bibliography{strings,refs}
}

\end{document}

%% file: sections/1_introduction.tex
\section{Introduction}
\label{sec:intro}

\vspace{-1mm}


Large audio–language models (LALMs) like SALMONN~\cite{tang2023salmonn}, Qwen-Audio-2~\cite{chu2024qwen2}, Audio Flamingo~\cite{kong2024audio}, and Step-Audio-2~\cite{wu2025step} have recently demonstrated impressive speech and audio understanding abilities.
However, benchmarks such as MMAU~\cite{sakshi2024mmau} reveal that perceptual errors still dominate over reasoning errors, underscoring that the bottleneck lies in the acoustic encoder.
Moreover, despite excelling at high-level semantic tasks, these LALMs still exhibit limited performance in voice-specific understanding~\cite{wang2024audiobench}, largely because upstream encoders, such as Whisper~\cite{radford2023robust} and BEATs~\cite{chen2022beats}, are primarily optimized for transcription rather than the nuanced characteristics of human voice.
This highlights the need for voice-aware encoders, which are critical for general-purpose speech understanding and personalized applications. 

\vspace{0.5mm}

In parallel, the voice representation community has developed specialized encoders. Models such as x-vector~\cite{snyder2018x} and ECAPA-TDNN~\cite{desplanques2020ecapa} remain strong baselines for speaker verification, while recent toolkit WeSpeaker~\cite{wang2023wespeaker} pushes further on identity robustness and generalizes to diverse tasks. On the paralinguistic side (e.g., emotion, age, gender), emotion2vec~\cite{ma2023emotion2vec} and Vesper~\cite{chen2024vesper} focus on affective and prosodic cues. However, these two directions are often pursued separately, leading to conflicting objectives: identity encoders suppress voice variability, whereas paralinguistic encoders emphasize it. Although some work explores unified representations~\cite{pappagari2020x, wang2024overview}, systematic training and evaluation across comprehensive tasks remain limited---not to mention the integration with LALMs.

Given that LALMs integrate acoustic encoders with language models, contrastive training strategies such as CLAP~\cite{elizalde2023clap} becomes a natural choice for aligning acoustic and linguistic spaces to facilitate understanding and reasoning. Recent efforts extend CLAP-style training to paralinguistic cues---including timbre, emotion, prosody~\cite{jing2024paraclap, dhamyal2024prompting, lin2024clap4emo, ra-clap}---but differences in datasets and evaluation protocols hinder systematic comparison. More critically, these works often overlook speaker identity, leaving open the question of how to design voice encoders that are language-aligned yet balanced across identity and paralinguistics.

Our work addresses these gaps through a systematic study of general-purpose voice encoders for speech and language understanding. We begin with task-specific initializations, then examine balance through multi-task training and CLAP fine-tuning, and finally evaluate trade-offs between identity and paralinguistics across diverse evaluations. This progressive analysis leads to our final encoder, \texttt{Auden-voice}.

\vspace{-5mm}
\input{tables/datasets}

%% file: tables/datasets.tex
\begin{table}[t]
\centering
\scriptsize
\renewcommand{\arraystretch}{0.44}   
\begin{threeparttable}
\caption{Datasets used for training and evaluation. 
LP = Linear Probing; ZS(C) = Zero-shot (Classification)}
\label{tab:datasets}
\begin{tabular}{p{10mm} p{38mm} p{10mm} p{9mm}}
\toprule
\multicolumn{4}{c}{{Training}} \\
\midrule
Task & Train Set(s) & \#Samples & Hours \\
\midrule
SID & VoxCeleb2\cite{chung2018voxceleb2}  & 974k & 2026 \\
Paraling & \makecell[l]{CREMA-D\cite{cao2014crema}, RAVDESS\cite{livingstone2018ryerson} \\IEMOCAP\cite{busso2008iemocap}, TESS\cite{SP2/E8H2MF_2020}}  & 18.3k& 20  \\
CLAP & \makecell[l]{ParaSpeechCaps \cite{diwan2025scaling}  (EARS\cite{richter2024ears}\\ EXPRESSO\cite{nguyen2023expresso}, Vox1\&2, Emilia\cite{he2024emilia})}  & \makecell[l]{base 111k\tnote{1}\\scaled 925k} & 2700\\
LLM-QA & \makecell[l]{CommonVoice\cite{ardila2019common}, IEMOCAP\\ MELD\cite{poria2018meld}, Vox2, CREMA-D \\ TESS, RAVDESS} & 1.76 M & 3250 \\
\midrule
\multicolumn{4}{c}{{Evaluation}} \\
\midrule
Task & Eval Set(s) & Eval & \#Samples \\
\midrule
SID & VoxCeleb2 & LP & 108k \\
Age & CREMA-D & LP/ZSC & 706  \\
Gender & CREMA-D, RAVDESS & LP/ZSC & 706, 136 \\
Emotion & CREMA-D, RAVDESS & LP/ZSC & 706, 136 \\
SV & VoxCeleb1-o\cite{Nagrani17} & ZS & 37k \\
SD/SC & VoxConverse\cite{Chung20} & ZS & 232 \\
Retrieval & ParaSpeechCaps-base & Retrieval & 568 \\
LLM-QA & \makecell[l]{AIR-Bench\cite{yang2024air}\tnote{2} (MELD\\IEMOCAP\tnote{3}, CommonVoice)} & QA & 5k \\
\bottomrule
\end{tabular}

\begin{tablenotes}[flushleft]
\scriptsize
\item[1] ParaSpeechCaps-train overlaps Vox1\&2 test; filtered in our CLAP training.
\item[2] AIR-Bench overlaps CommonVoice-train/dev; filtered in our LLM-QA training.
\item[3] IEMOCAP has no fixed split; minor leakage possible at init stage.
\end{tablenotes}
\vspace{-6mm}
\end{threeparttable}
\end{table}

%% file: sections/2_method.tex
\section{Method}
\label{sec:method}

\begin{figure*}[t]
\centerline{\includegraphics[width=0.98\linewidth,clip,trim=20 213 30 127]{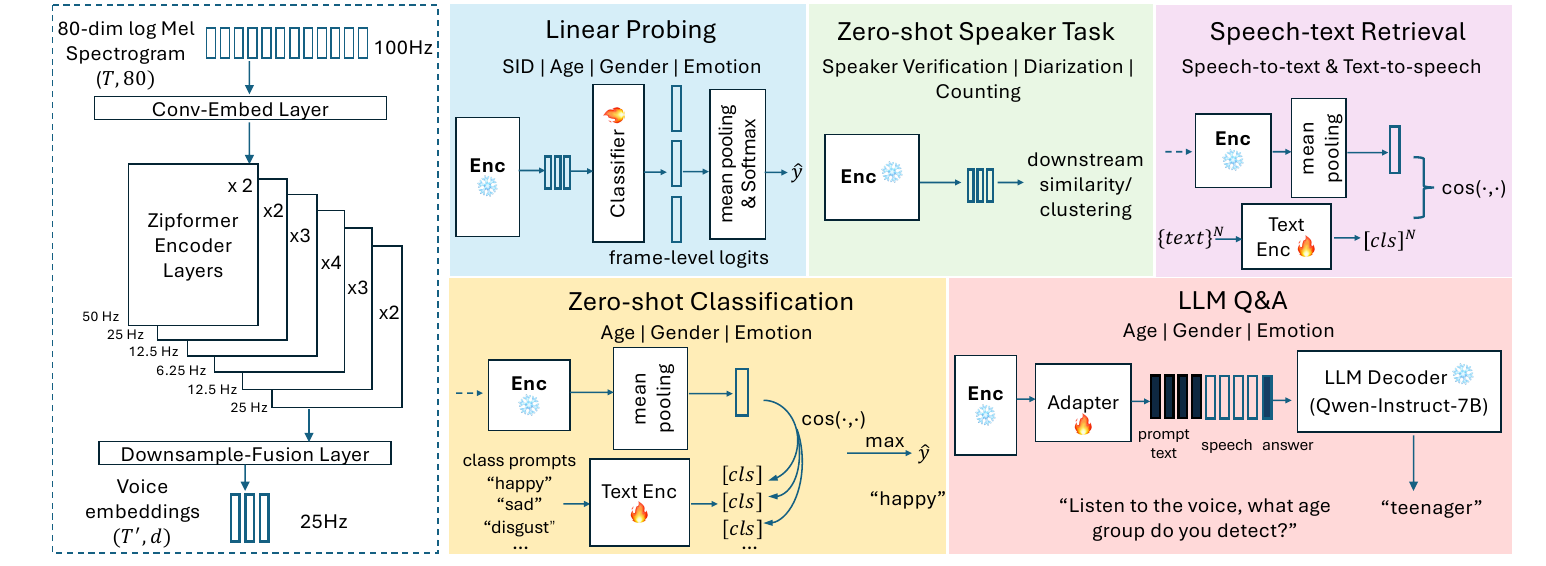}}
\vspace{-3mm}
\caption{Left: Zipformer encoder used across all settings to ensure fair evaluation. Right: Five evaluation setups covering both speech and speech-language understanding. Encoder is frozen; all heads are minimal to isolate voice representation quality.}
\vspace{-3mm}
\label{fig:figure}
\end{figure*}

\vspace{-2mm}
\subsection{Encoder Model}
\label{secsec:model}

We adopt Zipformer~\cite{yao2023zipformer} as the backbone. As shown in Figure~\ref{fig:figure} left, the model uses multi-rate Transformer layers to capture both fine-grained and long-range temporal patterns, followed by a downsample–fusion module that produces 25 Hz frame-level embeddings. To ensure fairness, all experiments use the same encoder (156M parameters) and always take the final-layer output. For tasks requiring utterance-level embeddings, we uniformly apply mean pooling, so that performance differences reflect encoder quality rather than task-specific aggregation, albeit sometimes at the cost of absolute accuracy. The resulting representation is 768-dimensional.


\vspace{-3mm}

\subsection{Training Strategies}
\label{secsec:training}

We progressively examine how supervision stages influence voice representation. Table~\ref{tab:datasets} summarizes the datasets.

(1) \textbf{Initialization.} 
We explore several supervision types: \textit{speaker identification} (SID) with cross-entropy (CE) and marginal loss, \textit{speaker verification} (SV) with generalized end-to-end loss~\cite{wan2018ge2e}, and paralinguistic classification jointly covering \textit{age},  \textit{gender}, and \textit{emotion}. All starts from an ASR-pretrained encoder, trained on large-scale in-house Chinese ASR data with the RNNT loss. This enables controlled analysis of the inductive biases introduced by each supervision. In practice, CE loss gave the most balanced initialization, so we report only these in later experiments.

(2) \textbf{Multi-task learning.} SID and the three paralinguistic tasks are trained jointly with balanced CE losses to encourage complementary representations.  When a sample lacks labels for certain tasks, pseudo labels are used.

(3) \textbf{CLAP fine-tuning.} On top of the above stages, we apply a speech–text contrastive loss that aligns paired voice and text embeddings in both directions, encouraging semantic grounding for zero-shot and paralinguistic generalization.

All encoders are trained with scaled Adam and Eden scheduling~\cite{yao2023zipformer} (lr=0.0045) on 32GB V100 GPUs. Audio is sampled at 16kHz with SpecAugment; no speed perturbation or additive noise is used. For fairness, all encoders and baselines share identical settings, including global max duration per batch whenever memory permits.

\vspace{-3mm}


%% file: tables/result_1.tex
\begin{table*}[!t]
\centering
\footnotesize                               
\setlength{\tabcolsep}{5pt}              
\renewcommand{\arraystretch}{0.42}       

\definecolor{lpHead}{RGB}{220,235,255}   
\definecolor{zsHead}{RGB}{220,245,225}   
\newcolumntype{Z}{>{\arraybackslash}p{6.5mm}} 
\newcolumntype{Q}{>{\arraybackslash}p{13mm}}  

\newcolumntype{L}{>{\columncolor{lpHead}\arraybackslash}p{6mm}} 
\newcolumntype{S}{>{\columncolor{zsHead}\arraybackslash}p{6.5mm}} 

\begin{threeparttable}
\caption{Linear probing (Acc↑) and speaker-related zero-shot results across initialization, supervision, and baselines. \textbf{LP Avg↑} is the mean of accuracy; \textbf{ZS Avg↑} is the negative mean of min-max normalized metrics.}
\label{tab:main_results_compact}

\begin{tabular}{@{}c Q Q Z Z Z Z Z Z L Z Z Z Z S@{}}

\toprule

\multirow{3}{*}[-0.5ex]{\textbf{Encoder\#}} &
\multirow{3}{*}[-0.5ex]{\textbf{Init.}} &
\multirow{3}{*}[-0.5ex]{\textbf{Supervision}} &
\multicolumn{7}{c}{\cellcolor{lpHead}\textbf{Linear probing}} &
\multicolumn{5}{c}{\cellcolor{zsHead}\textbf{Speaker Zero-shot}} \\
\cmidrule(lr{0.25em}){4-10} \cmidrule(l{0.25em}r{0.25em}){11-15}

& & &
\shortstack{\scriptsize\textbf{SID}\\ \tiny Vox2} &
\shortstack{\scriptsize\textbf{Age}\\{\tiny CREMA}} &
\shortstack{\scriptsize\textbf{Gender}\\{\tiny CREMA}} &
\shortstack{\scriptsize\textbf{Gender}\\{\tiny RAVDESS}} &
\shortstack{\scriptsize\textbf{Emo}\\{\tiny CREMA}} &
\shortstack{\scriptsize\textbf{Emo}\\{\tiny RAVDESS}} &
\shortstack{\scriptsize\textbf{LP}\\\textbf{Avg↑}} &
\shortstack{\textbf{SV}\\EER↓} &
\shortstack{\textbf{SD}\\DER↓} &
\shortstack{\textbf{SD}\\Conf↓} &
\shortstack{\textbf{Count}\\MAE↓} &
\shortstack{\textbf{ZS}\\\textbf{Avg↑}} \\  

\midrule
1.0 & --          & task-spec       & 84.8 & 92.3 & 92.2 & 99.4 & 65.4 & 81.5 & 85.9 & 8.5 & -- & -- & -- & -- \\
1.1 & --         & ASR        & 21.6 & 67.7 & 91.4 & 98.5 & 62.2 & 75.7 & 69.5 & 45.7 & 51.1 & 43.6 & 4.7 & 2.3 \\
1.2 & ASR          & SID       & \textbf{99.0} & 85.1 & 99.2 & 100 & 73.8 & 83.8 & \underline{90.2} & \textbf{2.3} & \textbf{14.2} & \textbf{6.8} & \underline{1.8} & \textbf{94.4} \\
1.3 & ASR          & Paraling  & 57.7 & \textbf{97.9} & \textbf{100} & 100 & \underline{79.8} & \textbf{94.1} & 88.3 & 37.1 & 50.0 & 42.5 & 4.5 & 9.8 \\
1.4 & ASR          & multi-task     & {95.3} & \underline{93.9} & \underline{99.7} & 100 & \textbf{84.0} & \underline{89.7} & \textbf{93.8} &\underline{3.8} & \underline{17.0} & \underline{9.5} & \textbf{1.6} & \underline{91.6} \\
\midrule
Whisper-medium\cite{radford2023robust} & --   & ASR     & 72.7 & 79.2 & 99.3 & 100 & 75.3 & {88.2} & 85.8 & 40.3 & 51.1 & 43.7 & 4.6 & 5.6 \\
wav2vec2.0-base\cite{baevski2020wav2vec} & SSL  & --      & 51.6 & 70.5 & 98.7 & 100 & 56.1 & 70.8 & 74.6 &  41.6 & 49.8 & 42.7 & 4.2 & 9.5 \\
emotion2vec\cite{ma2023emotion2vec} & SSL  & Emotion & --   & --   & --   & -- & --   & 82.9\tnote{*} & --  & 42.2 & 52.6 & 45.2 & 4.6 & 2.8 \\
Wespeaker\cite{wang2023wespeaker} & --   & SID     & \underline{96.2} & 83.9 & 98.4 & 100 & 70.2 & 87.5 & 89.4 & 0.8\tnote{*} & 11.3\tnote{*} & 3.8\tnote{*} & -- & {100} \\
\bottomrule
\end{tabular}

\begin{tablenotes}[flushleft]\scriptsize
\item[*] Reported in the original paper. Other results are from our framework with consistent settings (e.g., mean pooling) and training datasets for fair comparison.

\end{tablenotes}
\end{threeparttable}
\vspace{-3mm}
\end{table*}

%% file: sections/3_evaluation.tex
\subsection{Evaluation Setup}
\label{secsec:evaluation}



We freeze the encoder and evaluate its representation quality in five complementary setups as shown in Fig.~\ref{fig:figure}, covering both supervised and zero-shot scenarios.




\textbf{Linear probing.} For supervised classification tasks, a single linear layer is attached on top of frame-level encoder outputs $\mathbf{H} \in \mathbb{R}^{T' \times d}$. Then, frame-level logits are averaged across valid frames to obtain utterance-level predictions $\hat{y}$, trained with CE loss. We evaluate this setup on four tasks: SID (5994-way on VoxCeleb2-dev), age classification (four bins: teenager, yound adult, middle-aged adult, senior), binary gender classification, and emotion recognition with dataset-specific label sets such as 6-way for CREMA-D and 8-way for RAVDESS.

\textbf{Zero-shot speaker tasks.} We evaluate frozen utterance-level embeddings $\mathbf{v} \in \mathbb{R}^d$ (mean-pooled from $\mathbf{H}$) on three tasks without task-specific training. For SV, cosine similarity is used and equal error rate (EER) is reported on VoxCeleb1-o. For \textit{speaker diarization (SD)}, we adopt the PyAnnote~\cite{Plaquet23} pipeline3.1 with only the embedding model replaced, and evaluate on VoxConverse using diarization error rate (DER) and speaker confusion percentage. For \textit{speaker counting} (SC), the number of diarization clusters is compared with ground-truth counts via mean absolute error (MAE).

\textbf{Speech–text retrieval.} Motivated by applications such as text-guided target speaker extraction~\cite{huo2025beyond}, we evaluate the modality alignment between speech and text description via retrieval in both directions. The voice encoder is frozen, while a RoBERTa-base~\cite{DBLP:journals/corr/abs-1907-11692} text encoder is fine-tuned with the same CLAP-style contrastive loss used in Sec.~\ref{secsec:training}. Given paired embeddings $\{(\mathbf{v}_i, \mathbf{z}_i)\}_{i=1}^N$, where $\mathbf{v}_i$ is the pooled voice embedding and $\mathbf{z}_i$ is the \texttt{[CLS]} embedding from text description, similarity is measured by cosine distance:
\[
s(i,j) = \cos(\mathbf{v}_i, \mathbf{z}_j).
\]
Retrieval is evaluated by recall@1/5/10, i.e., whether the paired item appears in the top-k candidates, averaged over five 568-pair subsets for both directions.


\textbf{Zero-shot paralinguistic classification.} It allows immediate deployment on new classification tasks and unseen classes without additional training. Each voice embedding $\mathbf{v} \in \mathbb{R}^d$ is compared with prompt embeddings $\mathbf{z}_i \in \mathbb{R}^d$ from a fine-tuned text encoder, and classification is given by
\[
\hat{y} = \arg\max_i \cos(\mathbf{v}, \mathbf{z}_i).
\]
To mitigate sensitivity to prompt wording~\cite{dhamyal2024prompting,lin2024clap4emo}, we use the averaged text embedding from 10 natural language templates (e.g., ``The speaker sounds happy'').

\textbf{LLM-QA.} To test whether the encoder provides sufficient acoustic grounding for reasoning, we integrate it with a frozen LLM in paralinguistic question–answering (QA) tasks. The encoder outputs frame-level embeddings $\mathbf{H}$, which are jointly downsampled by a factor of 4 and projected into the LLM input embedding dimension through a lightweight adaptor consisting of two linear layers with a ReLU activation, following \cite{ma2024embarrassingly,mu2025efficient}. The adaptor’s output is then concatenated with the text embeddings of the questions and answers. At inference, given speech signal and a question prompt (e.g., ``Listen to the speech, what is the speaker's emotion?"), the system autoregressively generates a short text answer (e.g., ``angry'').



%% file: sections/4_results.tex
\section{Results}
\label{sec:results}

\subsection{Insights from Initialization and Supervision}

Table~\ref{tab:main_results_compact} presents an initial evaluation across voice-related tasks and highlights two fundamental observations. First, ASR-initialization (1.2–1.4) consistently outperforms training each task from scratch (1.0). Despite being optimized for semantic content, the ASR encoder (1.1) provides a strong inductive bias that benefits broader voice representation learning. Second, the choice of supervision yields task-specific advantages: SID supervision is more effective for speaker identity-related tasks, while paralinguistic supervision favors attributes such as age and emotion.

These observations suggest two natural directions for building more balanced voice encoders: (1) multi-task training that jointly optimizes SID and paralinguistic cues, and (2) CLAP fine-tuning to further strengthen paralinguistic capacity through speech-language alignment.




\subsection{Multi-task vs.\ CLAP for Balanced Representation}

The multi-task encoder (1.4) in Table~\ref{tab:main_results_compact} achieves the most balanced performance across speech understanding tasks. To further assess CLAP fine-tuning, Table~\ref{tab:retrieval_zs_cls} reports post-CLAP results with relative changes to their pre-CLAP counterparts. The most consistent gain from CLAP is speech--text retrieval, while its effects on other setups are mixed.

\begin{figure}
    \centering
    \includegraphics[width=0.9\columnwidth,clip,trim=105 210 255 130]{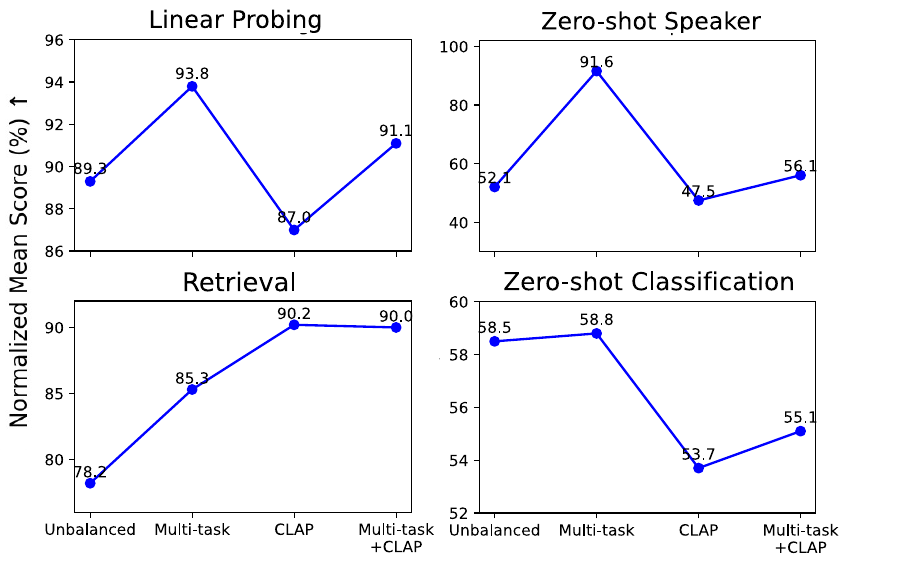}
    \vspace{-3mm}
    \caption{Task-specific effects of multi-task training and CLAP fine-tuning, comparing unbalanced baseline, multi-task, CLAP-tuned, and multi-task+CLAP encoders.}
    \vspace{-4mm}
    \label{fig:clap}
\end{figure}

For a clearer comparison, Fig.~\ref{fig:clap} summarizes averaged results for four representative encoders: the unbalanced single-supervision baseline (1.2 and 1.3 average), the multi-task encoder (1.4), the CLAP-tuned encoder (2.1), and the combined multi-task+CLAP encoder (2.4). Multi-task training consistently improves linear probing and speaker-related zero-shot tasks, whereas CLAP degrades them; in contrast, CLAP substantially outperforms multi-task on retrieval. Zero-shot classification sees modest gains from multi-task but declines after CLAP, contradicting our expectation that audio--language alignment would enhance paralinguistic capacity. In our earlier exploration, we also compared base vs.\ 10×-scaled CLAP training data and observed that scaling reliably boosts retrieval and trends toward recovering zero-shot classification ability, though still below pre-CLAP.

In summary, multi-task training provides more balanced improvements than CLAP, while the combined multi-task+CLAP encoder shows partial complementarity. We next examine their capacity for integration with LLMs, the ultimate goal of a general-purpose voice encoder.


\input{tables/result_2}


\subsection{LLM-QA with Frozen Voice Encoder}
\input{tables/llm-qa}

To assess which encoder provides stronger acoustic grounding for reasoning, we compare the multi-task (1.4) and the multi-task+CLAP (2.4) encoders on AIR-Bench~\cite{yang2024air} foundation tasks. Both the voice encoder and LLM are frozen, with only a lightweight adapter trained using template-based prompts and label answers; evaluation follows a multiple-choice format. Although training and evaluation differ in form, the adapter effectively bridges acoustic and textual modalities, showing strong instruction-following ability.

As shown in Table~\ref{tab:llmqa}, the multi-task encoder consistently outperforms the multi-task+CLAP encoder and is therefore adopted as the final \texttt{Auden-voice} encoder. Notably, QA performance correlates well with linear probing, i.e., the stronger probing results of the multi-task encoder translate into stronger LLM-QA grounding. However, adding CLAP on top of multi-task does not yield the expected gains in paralinguistic QA, likely because current CLAP training is still far from the scale of the vision–language community where scalability effects have emerged. Compared with end-to-end optimized and ASR--LLM cascade baselines, our modular approach achieves competitive results, underscoring the potential of \texttt{Auden-voice} as an LLM-integrable voice encoder for efficient speech understanding and reasoning.

\vspace{-2.9mm}

%% file: tables/result_2.tex
\definecolor{SecPurple}{RGB}{238,223,246}   
\definecolor{SecOrange}{RGB}{255,235,218}   
\definecolor{darkgreen}{RGB}{0,120,0}  
\definecolor{lpHead}{RGB}{220,235,255}   
\definecolor{zsHead}{RGB}{220,245,225}   

\newcolumntype{B}{>{\columncolor{SecPurple}}p{11mm}}   
\newcolumntype{G}{>{\columncolor{SecOrange}}p{11mm}}  

\newcolumntype{r}{>{\arraybackslash}p{10.5mm}}

\newcolumntype{Z}{>{\arraybackslash}p{11.9mm}} 
\newcolumntype{Q}{>{\arraybackslash}p{6.5mm}}  

\newcolumntype{L}{>{\columncolor{lpHead}\centering\arraybackslash}p{11mm}} 
\newcolumntype{S}{>{\columncolor{zsHead}\centering\arraybackslash}p{11mm}} 

\begin{table*}[!t]
\centering
\scriptsize
\setlength\tabcolsep{2pt}%
\renewcommand{\arraystretch}{0.9}%

\caption{
Speech–text retrieval and zero-shot classification results before and after CLAP fine-tuning. Absolute \textcolor{red}{increases} and \textcolor{darkgreen}{drops} are shown relative to pre-CLAP. The bottom block additionally reports LP and speaker-ZS of CLAP-fine-tuned encoders.
}
\label{tab:retrieval_zs_cls}
\begin{tabular}{@{} l l l Z r r r r r B r r r r r G @{}}
\toprule
\multirow{2}{*}{\textbf{Enc.\#}} &
\multirow{2}{*}{\textbf{Init.}} &
\multirow{2}{*}{\textbf{Sup.}} &
\multicolumn{7}{c}{\cellcolor{SecPurple}\textbf{Speech-text Retrieval}} &
\multicolumn{6}{c}{\cellcolor{SecOrange}\textbf{Zero-shot Classification (ZSC)}}\\
\cmidrule(lr){4-10}\cmidrule(lr){11-16}
& & &
\multicolumn{3}{c}{Speech-to-Text} &
\multicolumn{3}{c}{Text-to-Speech} &
\textbf{Retrieval} &
Age  &
Gender &
Gender  &
Emotion&
Emotion&
\textbf{ZSC}\\
& & &
R@1 & R@5 & R@10 & R@1 & R@5 & R@10 &  \textbf{Avg↑} &
\tiny{CREMA} & \tiny{CREMA} & \tiny{RAVDESS} & \tiny{CREMA} & \tiny{RAVDESS} &  \textbf{Avg↑} \\
\midrule
1.1 & -- & ASR & 35.9 & 68.4 & 77.8 & 32.6 & 64.5 & 75.3 & {59.1} & 21.3 & 76.8 & 67.7 & 29.2 & 20.6 & {43.1} \\
1.2 & ASR & SID & 62.7 & 96.1 & 98.1 & 61.0 & 96.3 & 98.7 & {85.5} & 22.0 & \underline{96.5} & \textbf{100} & 27.8 & 25.7 & {54.4} \\
1.3 & ASR & Paraling & 46.9 & 79.3 & 86.3 & 46.8 & 79.5 & 86.6 & {70.9} & 45.0 & 94.3 & 91.2 & \underline{36.4} & \underline{45.6} & \textbf{62.5} \\
1.4 & ASR & multi-task & 63.3 & 95.2 & 97.4 & 61.7 & 96.0 & 98.0 & {85.3} & 11.0 & \textbf{96.6} & 94.9 & \textbf{42.1} & \textbf{49.3} & \underline{58.8} \\

\midrule

2.1 & 1.1 & CLAP & \underline{71.9} \textcolor{red!50}{+36.0} & {97.8}  \textcolor{red!50}{+29.4} & {99.1}  \textcolor{red!50}{+21.3} & \textbf{74.2}  \textcolor{red!50}{+41.6} & \underline{98.8}  \textcolor{red!50}{+34.3} & \underline{99.5}  \textcolor{red!50}{+24.2} & \underline{90.2} \textcolor{red!50}{+31.1}  & 38.5  \textcolor{red!50}{+17.2} & 82.2  \textcolor{red!50}{+5.4} & 95.6  \textcolor{red!50}{+27.9} & 30.2  \textcolor{red!50}{+1.0} & 32.4  \textcolor{red!50}{+11.8} & {53.7} \textcolor{red!50}{+10.6} \\
2.2 & 1.2 & CLAP & \textbf{72.3} \textcolor{red!50}{+9.6} & \textbf{98.4} \textcolor{red!50}{+2.3} & \textbf{99.4} \textcolor{red!50}{+1.3} & \underline{73.9} \textcolor{red!50}{+12.9} & \textbf{98.9} \textcolor{red!50}{+2.6} & \textbf{99.6} \textcolor{red!50}{+0.9} & {\textbf{90.4} \textcolor{red!50}{+4.9}} & 23.8 \textcolor{red!50}{+1.8} & 94.8 \textcolor{darkgreen!90}{-1.7} & 98.5 \textcolor{darkgreen!90}{-1.5} & 22.4 \textcolor{darkgreen!90}{-5.4} & 26.5 \textcolor{red!50}{+0.8} & {53.2 \textcolor{darkgreen!90}{-1.2} } \\
2.3 & 1.3 & CLAP & 70.4 \textcolor{red!50}{+23.5} & 97.6 \textcolor{red!50}{+18.3} & 99.0 \textcolor{red!50}{+12.7} & 72.1 \textcolor{red!50}{+25.3} & 98.1 \textcolor{red!50}{+18.6} & 99.3 \textcolor{red!50}{+12.7} & 89.4 \textcolor{red!50}{+18.5} & 25.5 \textcolor{darkgreen!90}{-19.5} & 64.0 \textcolor{darkgreen!90}{-30.3} & \underline{99.3} \textcolor{red!50}{+8.1} & 35.0 \textcolor{darkgreen!90}{-1.4} & 40.4 \textcolor{darkgreen!90}{-5.2} & {52.8} \textcolor{darkgreen!90}{-9.7}\\
2.4 & 1.4 & CLAP & 71.3 \textcolor{red!50}{+8.0} & \underline{98.1} \textcolor{red!50}{+2.9} & \underline{99.3} \textcolor{red!50}{+1.9} & 73.2 \textcolor{red!50}{+11.5} & 98.6 \textcolor{red!50}{+2.6} & \underline{99.5} \textcolor{red!50}{+1.5} & {90.0} \textcolor{red!50}{+4.7} & 37.8 \textcolor{red!50}{+26.8} & 89.2 \textcolor{darkgreen!90}{-7.4}  & 91.2 \textcolor{darkgreen!90}{-3.7} & 28.5  \textcolor{darkgreen!90}{-13.6} & 28.7 \textcolor{darkgreen!90}{-20.6} & {55.1}  \textcolor{darkgreen!90}{-3.7} \\

\midrule




\multicolumn{3}{l}{Whisper-medium}   & 43.9 & 77.1 & 84.1 & 40.1 & 75.3 & 83.5 & {67.3} & 42.0 & 86.3 & 69.9 & 28.2 & 29.4 & {51.2} \\
\multicolumn{3}{l}{wav2vec2.0-base}  & 25.8 & 55.0 & 65.8 & 27.5 & 56.2 & 66.2 & {49.4} & \textbf{46.6} & 66.7 & 76.5 & 22.7 & 17.6 & {46.0} \\
\multicolumn{3}{l}{emotion2vec}  & 29.0 & 60.2 & 69.9 & 26.5 & 55.7 & 66.1 & {51.2} & \underline{46.0} & 74.7 & 75.0 & 23.1 & 17.6 & {47.3} \\
\multicolumn{3}{l}{Wespeaker}  & 47.5 & 81.1 & 87.5 & 46.7 & 82.7 & 88.8 & {72.4} & 28.6 & 64.3 & 98.5 & 24.4 & 17.7 & {46.7} \\
\end{tabular}

\begin{tabular}{@{}Q Q Q Z Z Z Z Z Z L Z Z Z Z S@{}}
\toprule
\multirow{3}{*}[-0.5ex]{\textbf{Enc.\#}} &
\multirow{3}{*}[-0.5ex]{\textbf{Init.}} &
\multirow{3}{*}[-0.5ex]{\textbf{Sup.}} &
\multicolumn{7}{c}{\cellcolor{lpHead}\textbf{Linear probing (LP)}} &
\multicolumn{5}{c}{\cellcolor{zsHead}\textbf{Speaker Zero-shot}} \\
\cmidrule(lr{0.25em}){4-10} \cmidrule(l{0.25em}r{0.25em}){11-15}
& & &
\shortstack{\textbf{SID}\\Vox2} &
\shortstack{\textbf{Age}\\{\fontsize{6pt}{7pt}\selectfont CREMA}} &
\shortstack{\textbf{Gender}\\{\fontsize{6pt}{7pt}\selectfont CREMA}} &
\shortstack{\textbf{Gender}\\{\fontsize{6pt}{7pt}\selectfont RAVDESS}} &
\shortstack{\textbf{Emotion}\\{\fontsize{6pt}{7pt}\selectfont CREMA}} &
\shortstack{\textbf{Emotion}\\{\fontsize{6pt}{7pt}\selectfont RAVDESS}} &
\shortstack{\textbf{LP}\\\textbf{Avg↑}} &
\shortstack{\textbf{SV}\\EER↓} &
\shortstack{\textbf{SD}\\DER↓} &
\shortstack{\textbf{SD}\\Conf↓} &
\shortstack{\textbf{Count}\\MAE↓} &
\shortstack{\textbf{ZS}\\\textbf{Avg↑}} \\  
\midrule
2.1 & 1.1        & CLAP        & 83.0 \textcolor{red!50}{+61.4} & 77.9 \textcolor{red!50}{+10.2} & 99.6  \textcolor{red!50}{+8.2} & 100  \textcolor{red!50}{+1.5} & 72.7  \textcolor{red!50}{+10.5}  & 89.0  \textcolor{red!50}{+13.3} & 87.0 \textcolor{red!50}{+17.5} & 10.2 \textcolor{red!50}{+35.5} & 35.5 \textcolor{red!50}{+15.6} & 28.0  \textcolor{red!50}{+15.6} & 3.8 \textcolor{red!50}{+0.9} & 47.5 \textcolor{red!50}{+45.2} \\
2.2 & 1.2     & CLAP & 95.9 \textcolor{darkgreen!90}{-3.1} & 82.6 \textcolor{darkgreen!90}{-2.5} & 99.0 \textcolor{darkgreen!90}{-0.2} & 100 \textcolor{red!50}{+0.0} & 70.7 \textcolor{darkgreen!90}{-3.1} & 83.1 \textcolor{darkgreen!90}{-0.7} & 88.6 \textcolor{darkgreen!90}{-1.6} & 9.8 \textcolor{darkgreen!90}{-7.5} & 26.5 \textcolor{darkgreen!90}{-12.3} & 19.0 \textcolor{darkgreen!90}{-12.2} & 2.8 \textcolor{darkgreen!90}{-1.0} &66.9 \textcolor{darkgreen!90}{-27.5}\\
2.3 & 1.3 & CLAP & 88.4 \textcolor{red!50}{+30.7} & 89.8 \textcolor{darkgreen!90}{-8.1} & 99.9 \textcolor{darkgreen!90}{-0.1} & 100 \textcolor{red!50}{+0.0} & 77.5 \textcolor{darkgreen!90}{-2.3} & 89.0 \textcolor{darkgreen!90}{-5.1} & 90.8 \textcolor{red!50}{+2.5} & 10.6  \textcolor{red!50}{+26.5} & 37.8  \textcolor{red!50}{+12.2} & 30.4  \textcolor{red!50}{+12.1} & 3.9  \textcolor{red!50}{+0.6} & 43.8 \textcolor{red!50}{+34.0} \\
2.4 & 1.4           & CLAP     & 90.2 \textcolor{darkgreen!90}{-5.1} & 86.5 \textcolor{darkgreen!90}{-7.4} & 99.6 \textcolor{darkgreen!90}{-0.1} & 100 \textcolor{red!50}{+0.0} & 76.8 \textcolor{darkgreen!90}{-7.2} & {93.4} \textcolor{red!50}{+3.7} & 91.1 \textcolor{darkgreen!90}{-2.7} & 10.9 \textcolor{darkgreen!90}{-7.1} & 31.6 \textcolor{darkgreen!90}{-14.6} & 24.1 \textcolor{darkgreen!90}{-14.6} & 3.3 \textcolor{darkgreen!90}{-1.7} &56.1 \textcolor{darkgreen!90}{-35.5} \\
\bottomrule
\end{tabular}

\vspace{-2mm}
\end{table*}

%% file: tables/llm-qa.tex
\begin{table}[t]
\centering
\scriptsize
\setlength\tabcolsep{3pt} 
\caption{LLM-QA acc (\%) on AIR-Bench foundation benchmark. Top three rows freeze the voice encoder and the LLM (Qwen2.5-7B-Inst~\cite{qwen2.5}), training only a lightweight adapter.}

\label{tab:llmqa}
\resizebox{\columnwidth}{!}{%
\begin{tabular}{lccccc}
\toprule
\multirow{2}{*}{\textbf{System}} 
& \multicolumn{2}{c}{\textbf{Emotion}} 
& \multicolumn{2}{c}{\textbf{Gender}}
& \textbf{Age} \\
\cmidrule(lr){2-3}\cmidrule(lr){4-5}
& \textbf{MELD$^*$} & \textbf{IEMO} & \textbf{MELD} & \textbf{CV} & \textbf{CV} \\
\midrule
\makecell[l]{\textbf{Enc 1.4: multi-task}}      & 27.2  & \textbf{84.7} & \textbf{81.6} & \textbf{93.2} & 58.3 \\
\makecell[l]{Enc 2.4: multi-task+CLAP}      & {22.3} & 43.6 & 76.2 & 87.3 & \textbf{66.2} \\
\makecell[l]{Whisper + Qwen-Inst-7B}        & 42.2 & 27.5 & 47.6 & 52.2 & 65.3 \\
\makecell[l]{Qwen-Audio (end-to-end)}                                     & \multicolumn{2}{c}{43.2} & \multicolumn{2}{c}{67.2} & 36.0 \\
\makecell[l]{Whisper $\rightarrow$ GPT-4 (cascade)}                        & \multicolumn{2}{c}{59.5} & \multicolumn{2}{c}{21.9} & 41.1 \\
\bottomrule
\end{tabular}}

\begin{tablenotes}[flushleft]\scriptsize
\item[*] * Performance shows high variance across encoders despite similar results on other tasks. \\ This may due to the lack of diversity in LLM-QA training data.
\end{tablenotes}

\vspace{-4mm}

\end{table}

%% file: sections/5_discussion.tex
\section{Conclusion}
\label{sec:discussion}

\vspace{-2mm}

Our results highlight three key findings. First, ASR initialization provides a strong foundation, and multi-task supervision yields the most balanced voice representations across identity and paralinguistic cues. Second, while CLAP fine-tuning consistently enhances speech–text retrieval, it degrades other tasks, with scaling showing only a partial trend toward recovery. Third, LLM-QA performance correlates well with linear probing, and the multi-task encoder proves most effective for grounding LLM reasoning. 
Taken together, these insights lead to our final \texttt{Auden-voice} encoder, which balances identity and paralinguistic cues while integrating competitively with LLMs toward general-purpose voice encoder.

%% file: refs.bib
@article{yao2023zipformer,
  title={Zipformer: A faster and better encoder for automatic speech recognition},
  author={Yao, Z. and Guo, L. and Yang, X. and Kang, W. and Kuang, F. and Yang, Y. and Jin, Z. and Lin, L. and Povey, D.},
  journal={arXiv preprint arXiv:2310.11230},
  year={2023}
}

@inproceedings{wang2023wespeaker,
  title={Wespeaker: A research and production oriented speaker embedding learning toolkit},
  author={Wang, H. and Liang, C. and Wang, S. and Chen, Z. and Zhang, B. and Xiang, X. and Deng, Y. and Qian, Y.},
  booktitle={Proc. ICASSP 2023},
  year={2023}
}

@inproceedings{huo2025beyond,
  title={Beyond speaker identity: Text guided target speech extraction},
  author={Huo, M. and Jain, A. and Huynh, C. P. and Kong, F. and Wang, P. and Liu, Z. and Bhat, V.},
  booktitle={Proc. ICASSP 2025},
  year={2025}
}

@article{chung2018voxceleb2,
  title={{VoxCeleb2}: Deep speaker recognition},
  author={Chung, J. S. and Nagrani, A. and Zisserman, A.},
  journal={arXiv preprint arXiv:1806.05622},
  year={2018}
}

@inproceedings{Nagrani17,
  author={Nagrani, A. and Chung, J. S. and Zisserman, A.},
  title={{VoxCeleb}: A large-scale speaker identification dataset},
  booktitle={Proc. INTERSPEECH 2017},
  year={2017}
}

@article{cao2014crema,
  title={{CREMA-D}: Crowd-sourced emotional multimodal actors dataset},
  author={Cao, H. and Cooper, D. G. and Keutmann, M. K. and Gur, R. C. and Nenkova, A. and Verma, R.},
  journal={IEEE Trans. Affective Comput.},
  pages={377--390},
  year={2014}
}

@article{livingstone2018ryerson,
  title={The Ryerson Audio-Visual Database of Emotional Speech and Song {(RAVDESS)}: A dynamic, multimodal set of facial and vocal expressions in North American English},
  author={Livingstone, S. R. and Russo, F. A.},
  journal={PLOS One},
  volume={13},
  number={5},
  pages={e0196391},
  year={2018}
}

@inproceedings{Chung20,
  author={Chung, J. S. and Huh, J. and Nagrani, A. and Afouras, T. and Zisserman, A.},
  title={Spot the conversation: Speaker diarisation in the wild},
  booktitle={Proc. INTERSPEECH 2020},
  year={2020}
}

@article{diwan2025scaling,
  title={Scaling rich style-prompted text-to-speech datasets},
  author={Diwan, A. and Zheng, Z. and Harwath, D. and Choi, E.},
  journal={arXiv preprint arXiv:2503.04713},
  year={2025}
}

@article{nguyen2023expresso,
  title={Expresso: A benchmark and analysis of discrete expressive speech resynthesis},
  author={Nguyen, T. A. and Hsu, W.-N. and d'Avirro, A. and Shi, B. and Gat, I. and Fazel-Zarani, M. and Remez, T. and Copet, J. and Synnaeve, G. and Hassid, M. et al.},
  journal={arXiv preprint arXiv:2308.05725},
  year={2023}
}

@article{richter2024ears,
  title={{EARS}: An anechoic fullband speech dataset benchmarked for speech enhancement and dereverberation},
  author={Richter, J. and Wu, Y.-C. and Krenn, S. and Welker, S. and Lay, B. and Watanabe, S. and Richard, A. and Gerkmann, T.},
  journal={arXiv preprint arXiv:2406.06185},
  year={2024}
}

@inproceedings{he2024emilia,
  title={Emilia: An extensive, multilingual, and diverse speech dataset for large-scale speech generation},
  author={He, H. and Shang, Z. and Wang, C. and Li, X. and Gu, Y. and Hua, H. and Liu, L. and Yang, C. and Li, J. and Shi, P. et al.},
  booktitle={Proc. SLT 2024},
  pages={885--890},
  year={2024}
}

@article{poria2018meld,
  title={{MELD}: A multimodal multi-party dataset for emotion recognition in conversations},
  author={Poria, S. and Hazarika, D. and Majumder, N. and Naik, G. and Cambria, E. and Mihalcea, R.},
  journal={arXiv preprint arXiv:1810.02508},
  year={2018}
}

@article{busso2008iemocap,
  title={{IEMOCAP}: Interactive emotional dyadic motion capture database},
  author={Busso, C. and Bulut, M. and Lee, C.-C. and Kazemzadeh, A. and Mower, E. and Kim, S. and Chang, J. N. and Lee, S. and Narayanan, S. S.},
  journal={Lang. Resour. Eval.},
  volume={42},
  number={4},
  pages={335--359},
  year={2008}
}

@article{ma2024embarrassingly,
  title={An embarrassingly simple approach for llm with strong asr capacity},
  author={Ma, Z. and Yang, G. and Yang, Y. and Gao, Z. and Wang, J. and Du, Z. and Yu, F. and Chen, Q. and Zheng, S. and Zhang, S. and others},
  journal={arXiv preprint arXiv:2402.08846},
  year={2024}
}

@article{mu2025efficient,
  title={Efficient Scaling for {LLM}-based {ASR}},
  author={Mu, B. and Shao, Y. and Wei, K. and Yu, D. and Xie, L.},
  journal={arXiv preprint arXiv:2508.04096},
  year={2025}
}

@article{ardila2019common,
  title={Common Voice: A massively multilingual speech corpus},
  author={Ardila, R. and Branson, M. and Davis, K. and Henretty, M. and Kohler, M. and Meyer, J. and Morais, R. and Saunders, L. and Tyers, F. M. and Weber, G.},
  journal={arXiv preprint arXiv:1912.06670},
  year={2019}
}

@misc{SP2/E8H2MF_2020,
  author={Pichora-Fuller, M. K. and Dupuis, K.},
  title={Toronto emotional speech set ({TESS})},
  year={2020},
  howpublished={https://doi.org/10.5683/SP2/E8H2MF},
}

@article{yang2024air,
  title={{AIR-Bench}: Benchmarking large audio-language models via generative comprehension},
  author={Yang, Q. and Xu, J. and Liu, W. and Chu, Y. and Jiang, Z. and Zhou, X. et al.},
  journal={arXiv preprint arXiv:2402.07729},
  year={2024}
}

@article{chu2024qwen2,
  title={Qwen2-audio technical report},
  author={Chu, Y. and Xu, J. and Yang, Q. and Wei, H. and Wei, X. and Guo, Z. et al.},
  journal={arXiv preprint arXiv:2407.10759},
  year={2024}
}

@article{tang2023salmonn,
  title={{SALMONN}: Towards generic hearing abilities for large language models},
  author={Tang, C. and Yu, W. and Sun, G. and Chen, X. and Tan, T. and Li, W. and Lu, L. and Ma, Z. and Zhang, C.},
  journal={arXiv preprint arXiv:2310.13289},
  year={2023}
}

@article{kong2024audio,
  title={Audio Flamingo: A novel audio language model with few-shot learning and dialogue abilities},
  author={Kong, Z. and Goel, A. and Badlani, R. and Ping, W. and Valle, R. and Catanzaro, B.},
  journal={arXiv preprint arXiv:2402.01831},
  year={2024}
}

@article{wu2025step,
  title={{Step-Audio 2} technical report},
  author={Wu, B. and Yan, C. and Hu, C. and Yi, C. and Feng, C. and Tian, F. et al.},
  journal={arXiv preprint arXiv:2507.16632},
  year={2025}
}

@article{sakshi2024mmau,
  title={{MMAU}: A massive multi-task audio understanding and reasoning benchmark},
  author={Sakshi, S. and Tyagi, U. and Kumar, S. and Seth, A. and Selvakumar, R. and Nieto, O. et al.},
  journal={arXiv preprint arXiv:2410.19168},
  year={2024}
}

@inproceedings{radford2023robust,
  title={Robust speech recognition via large-scale weak supervision},
  author={Radford, A. and Kim, J. W. and Xu, T. and Brockman, G. and McLeavey, C. and Sutskever, I.},
  booktitle={Proc. ICML 2023},
  pages={28492--28518},
  year={2023}
}

@article{chen2022beats,
  title={{BEATS}: Audio pre-training with acoustic tokenizers},
  author={Chen, S. and Wu, Y. and Wang, C. and Liu, S. and Tompkins, D. and Chen, Z. and Wei, F.},
  journal={arXiv preprint arXiv:2212.09058},
  year={2022}
}

@article{desplanques2020ecapa,
  title={{ECAPA-TDNN}: Emphasized channel attention, propagation and aggregation in TDNN based speaker verification},
  author={Desplanques, B. and Thienpondt, J. and Demuynck, K.},
  journal={arXiv preprint arXiv:2005.07143},
  year={2020}
}

@inproceedings{snyder2018x,
  title={X-vectors: Robust {DNN} embeddings for speaker recognition},
  author={Snyder, D. and Garcia-Romero, D. and Sell, G. and Povey, D. and Khudanpur, S.},
  booktitle={Proc. ICASSP 2018},
  pages={5329--5333},
  year={2018}
}

@article{ma2023emotion2vec,
  title={Emotion2vec: Self-supervised pre-training for speech emotion representation},
  author={Ma, Z. and Zheng, Z. and Ye, J. and Li, J. and Gao, Z. and Zhang, S. and Chen, X.},
  journal={arXiv preprint arXiv:2312.15185},
  year={2023}
}

@article{chen2024vesper,
  title={Vesper: A compact and effective pretrained model for speech emotion recognition},
  author={Chen, W. and Xing, X. and Chen, P. and Xu, X.},
  journal={IEEE Trans. Affective Comput.},
  volume={15},
  number={3},
  pages={1711--1724},
  year={2024}
}

@inproceedings{elizalde2023clap,
  title={{CLAP}: Learning audio concepts from natural language supervision},
  author={Elizalde, B. and Deshmukh, S. and Al Ismail, M. and Wang, H.},
  booktitle={Proc. ICASSP 2023},
  year={2023}
}

@article{wang2024audiobench,
  title={{AudioBench}: A universal benchmark for audio large language models},
  author={Wang, B. and Zou, X. and Lin, G. and Sun, S. and Liu, Z. and Zhang, W. and Liu, Z. and Aw, A. and Chen, N. F.},
  journal={arXiv preprint arXiv:2406.16020},
  year={2024}
}

@inproceedings{pappagari2020x,
  title={X-vectors meet emotions: A study on dependencies between emotion and speaker recognition},
  author={Pappagari, R. and Wang, T. and Villalba, J. and Chen, N. and Dehak, N.},
  booktitle={Proc. ICASSP 2020},
  year={2020}
}

@article{wang2024overview,
  title={Overview of speaker modeling and its applications},
  author={Wang, S. and Chen, Z. and Lee, K. A. and Qian, Y. and Li, H.},
  journal={IEEE/ACM Trans. Audio, Speech, Lang. Process.},
  year={2024}
}

@article{jing2024paraclap,
  title={{ParaCLAP}: Towards a general language-audio model for computational paralinguistic tasks},
  author={Jing, X. and Triantafyllopoulos, A. and Schuller, B.},
  journal={arXiv preprint arXiv:2406.07203},
  year={2024}
}

@inproceedings{dhamyal2024prompting,
  title={Prompting audios using acoustic properties for emotion representation},
  author={Dhamyal, H. and Elizalde, B. and Deshmukh, S. and Wang, H. and Raj, B. and Singh, R.},
  booktitle={Proc. ICASSP 2024},
  year={2024}
}

@inproceedings{lin2024clap4emo,
  author={Lin, W.-C. and Ghaffarzadegan, S. and Bondi, L. and Kumar, A. and Das, S. and Wu, H.-H.},
  title={{CLAP4Emo}: {ChatGPT}-assisted speech emotion retrieval with natural language supervision},
  booktitle={Proc. ICASSP 2024},
  year={2024},
}

@article{ra-clap,
  title={{RA-CLAP}: Relation-augmented emotional speaking style contrastive language-audio pretraining for speech retrieval},
  author={Sun, H. and Tian, J. and Zhou, J. and Wang, H. and He, J. and Zhao, S. and Kong, X. et al.},
  journal={arXiv preprint arXiv:2505.19437},
  year={2025}
}

@inproceedings{wan2018ge2e,
  author    = {L. Wan and Q. Wang and A. Papir and I. L. Moreno},
  title     = {Generalized End-to-End Loss for Speaker Verification},
  booktitle = {Proc. ICASSP 2018},
  year      = {2018}
}

@inproceedings{baevski2020wav2vec,
  author    = {A. Baevski and Y. Zhou and A. Mohamed and M. Auli},
  title     = {wav2vec 2.0: A Framework for Self-Supervised Learning of Speech Representations},
  booktitle = {Proc. NeurIPS 2020},
  pages     = {12449--12460},
  year      = {2020}
}

@misc{qwen2.5,
    title = {Qwen2.5: A Party of Foundation Models},
    url = {https://qwenlm.github.io/blog/qwen2.5/},
    author = {Qwen Team},
    year = {2024}
}

@article{DBLP:journals/corr/abs-1907-11692,
  author    = {Liu, L. and
                Ott, M. and
                Goyal, N. and
                Du, J. and
                Joshi, M. and
                Chen, D. and
                Levy, O. and
                Lewis, M. and
                Zettlemoyer, L. and
                Stoyanov,V.},
  title     = {RoBERTa: {A} Robustly Optimized {BERT} Pretraining Approach},
  year      = {2019}
}

@inproceedings{Plaquet23,
  author={Plaquet, A. and Bredin, H.},
  title={{Powerset multi-class cross entropy loss for neural speaker diarization}},
  year=2023,
  booktitle={Proc. INTERSPEECH 2023},
}
